# Highly luminescent *a*-SiO$_x$<Er>/SiO$_2$/Si multilayer structure

Rossano Lang, David S. L. Figueira, Felipe Vallini and Newton C. Frateschi

Device Research Laboratory, Applied Physics Department, "Gleb Wataghin" Physics Institute, University of Campinas - UNICAMP 13083-859 Campinas, SP, Brazil

**Abstract:** We have fabricated highly-luminescent samples with erbium-doped amorphous silicon sub-oxide (a-SiO$_x$<Er>) layers on SiO$_2$/Si substrates. The layers are designed to provide a resonance with large modal overlap with the active material and with low quality factor (Q-factor) at 1540 nm. Also, the structure has higher Q-factor resonances in the wavelength range between 600 - 1200 nm. Within this range, strong light emission from a-SiOx defect-related radiative centers and emission from the Er$^{3+}$ optical transition $^4I_{11/2}$ - $^4I_{15/2}$ (980 nm) are observed. A two-fold improvement in photoluminescence (PL) intensity is achieved in the wavelength range between 800 - 1000 nm. The PL intensity in the wavelength range between 1400 - 1700 nm (region of Er$^{3+}$ $^4I_{13/2}$ - $^4I_{15/2}$ transition) is increased four times. This later higher intensity enhancement is apparently caused by optical pumping at 980 nm, higher Q-factor, with subsequent emission from the $^4I_{13/2}$ level in the low Q resonance at 1540 nm. Further five times emission enhancement is obtained after optimized temperature annealing. The temperature-induced quenching in the PL intensity indicates distinct deactivation energies related to different types of Er centers which are more or less coupled to defects depending on the thermal treatment temperature.

## 1. Introduction

Numerous attempts towards obtaining gain or employing Si as an active medium for direct application in optoelectronics devices have been reported [1-3]. However, this task has been shown to be difficult and several alternatives have been proposed [4-6]. In this context, Erbium-doped silicon-based materials are attractive for telecom applications due to the luminescence originated from the transition between the Er$^{3+}$ two lowest spin-orbit levels ($^4I_{13/2}$ - $^4I_{15/2}$) that occurs at 1540 nm, the C-band center.

In an amorphous Si matrix, the Er$^{3+}$ excitation mechanism is associated to a defect related Auger quasiresonant process (DRAE) involving dangling bonds states and the Er$^{3+}$ ground state [7] and/or a resonant dipole-dipole interaction originated by the non-radiative recombination of the electron-hole pairs in the dangling bonds [8]. In both cases, electrons of the 4f shell are excited from the $^4I_{15/2}$ to the $^4I_{13/2}$ level. It has also been suggested a direct pumping of the $^4I_{11/2}$ level by either tail-to-tail states recombination, a similar mechanism as above, or by up conversion caused by Er-Er interaction [9]. However, a much more efficient approach to populate the $^4I_{11/2}$ level is by external optical pump at 980 nm such as in Er-doped fiber amplifiers (EDFA) [10]. The electrons decay non-radiatively to the $^4I_{13/2}$ level and provide gain to the $^4I_{13/2}$ - $^4I_{15/2}$ transition. Another way, less frequently used, is via the $^4I_{15/2}$ - $^4I_{9/2}$ transition at 807 nm [11]. This transition has a small optical absorption cross section (typically on the order of 10$^{-21}$ - 10$^{-20}$ cm$^2$, depending on the host material) [12]. For this reason, there is significant interest in sensitizing Er$^{3+}$ ions by adding strongly absorbing elements that can transfer energy efficiently to Er [13]. From this point of view, some reports have shown evidence of the direct energy transfer from silicon nanocrystals (Si-NC's) to Er$^{3+}$ through non-radiative recombination of excitons [14,15]. However, Kuritsyn *et al*. argued that Si-NC's embedded in silicon oxide are not specific in any respect when compared to other types of defects due to excess silicon and a noteworthy defect-mediated and resonant optical excitation of Er$^{3+}$ was verified [16]. We have demonstrated an enhancement of the emission at 1540 nm when erbium-doped amorphous silicon sub-oxide (*a*-SiO$_x$<Er>) layers with Si-NC's were properly placed in vertical low Q-factor mutilayer structure [17]. However, high temperature annealing such as 1000 °C required for the Si-NC's formation, have given rise to a strong deterioration in the emission with respect to the material annealed at lower temperatures.

In this work, instead of employing Si-NC's for electronic excitation transfer to the Er ions, we have optimized the annealing temperature to enhance the emission in a broad wavelength range (800 - 1000 nm). This emission provides the optical pumping at 980 nm which is enhanced by a resonance near this wavelength. The pumping at 980 nm leads to emission at 1540 nm that is extracted from the structure through a low Q-factor, hence, high external quantum efficiency resonance. Essentially this creates a cross pumping mechanism. The emission in the 800 - 1000 nm range is mostly provided by optically active defect centers from the *a*-SiO$_x$ matrix. It is important to observe that there is a considerable spectral overlap of the intrinsic *a*-SiO$_x$ optical emission with the transition energies from the Er$^{3+}$ $^4I_{15/2}$ level to the $^4I_{9/2}$ (1.536 eV - 807 nm) and $^4I_{11/2}$ (1.265 eV - 980 nm) levels. The vertical resonant structure consists of an *a*-SiO$_x$<Er> layer deposited on a SiO$_2$ thermal oxide. The thicknesses of both layers are chosen to provide the largest overlap of the optical mode and the Er$^{3+}$ doped material at the wavelength of 1540 nm. This was done by optimizing the electric field intensity resulting from the counter-propagating fields after a single reflection at the $a$-SiO$_x$<Er>/air - ($E_{\text{air}}$) and the $a$-SiO$_x$<Er>/SiO$_2$ - ($E_{\text{SiO}_2}$) interfaces in all points of the $a$-SiO$_x$<Er> layer. In other words, we have optimized the integral:

$$\Gamma = (1/L)\int_0^L \left| E_{\text{air}}(x) + E_{\text{SiO}_2}(x) \right|^2 dx \qquad (1)$$

at the 1540 nm wavelength, where $L$ is cavity length. This approach is valid for small Q structures and optimizes the modal effective volume within the $a$-SiO$_x$<Er>. However, as it will be shown later, we also need to consider the higher Q modes that exist in the 800 - 1000 nm wavelength range. We show that the resonances within this range enhances the optical emission by over 2 times and, as a consequence, they enhance the emission at 1540 nm by over 4 times due to the cross pumping mechanism. After annealing optimization, we obtain a further 5 times increase in the emission at 1540 nm as compared to the as-deposited on SiO$_2$ sample.

## 2. Experimental Details

As the starting material we used an $n$-type Czochralski Si (001) wafer (thickness 500 μm). A 530 nm SiO$_2$ layer was obtained by wet-oxidation at 1000 °C. Subsequently, a 600 nm thick $a$-SiO$_x$<Er> film was growth by reactive co-sputtering deposition. This heterostructure results in the largest integral Γ at 1540 nm. Rutherford Backscattering Spectrometry (RBS) indicated an Er concentration of ~ 0.01 at.% and an O concentration of ~ 7.8 at.%. A 600 nm thick $a$-SiO$_x$<Er> film was also deposited directly on Si to verify the resonance effect. After deposition, some samples were thermally annealed in a N$_2$ atmosphere (3.0 l/min flux) at 400 °C and at 1150 °C for 1 hour in order to investigate the light emission evolution. Photoluminescence spectroscopy (PL) was carried out at room temperature and at low temperatures in a continuous flow variable-temperature He cryostat. We used a 532 nm line of a Solid State Laser as photoexcitation source. The spot area was ~ 1 mm$^2$. The PL was dispersed by a single-grating monochromator (SPEX 0.5 m focal length with 600 l/mm grating and 32 Å/mm resolution) and detected by a photomultiplier S1 Hamamatsu (600 - 1200 nm spectral region) and by a Ge $p$-$i$-$n$ photodiode North Coast (1400 - 1700 nm wavelength range). Both detectors were cooled by liquid nitrogen.

## 3. Results and Discussion

Figure 1 shows the PL spectra in the abovementioned wavelength ranges for $a$-SiO$_x$<Er> layers as-deposited on 530 nm SiO$_2$ and directly on Si. These measurements were performed with an estimated excitation power density of 25 mW/mm$^2$. The PL spectra were normalized to the maximum intensity at both ranges. The emission intensity is clearly enhanced for the samples with the SiO$_2$ layer. More specifically, a two-fold enhancement is observed at in the 600 - 1200 nm range while a four-fold enhancement is observed at ~ 1540 nm. Figure 1 also shows the calculated optical field intensity after infinite roundtrips as a function of wavelength at any arbitrary point inside the $a$-SiO$_x$<Er> layer for samples with and without the SiO$_2$ film. The sample with the SiO$_2$ layer has resonances near 800 nm and 950 nm with Q of ~ 13 and ~ 17, respectively. A resonance with small Q, Q ~ 3.5, is observed at 1535 nm. The sample without the SiO$_2$ layer has only small Q resonances, Q ~ 3.5, near 760 nm and 940 nm.

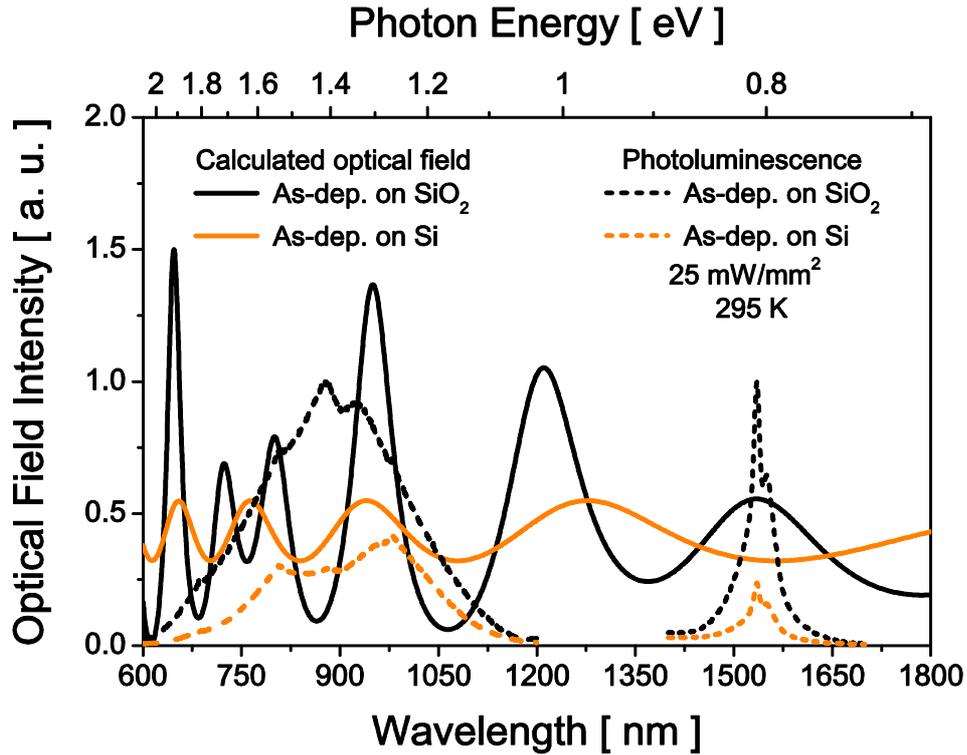

**Figure 1**. PL spectra at 295 K in the 600 - 1200 nm and 1400 - 1700 nm wavelength ranges of $a$-SiO$_x$<Er> layers as-deposited on 530 nm SiO$_2$ and on S in contrast with the calculated optical field intensity inside the $a$-SiO$_x$<Er> layer for structures with and without the SiO$_2$ film.

A close inspection of the optical emission in the 600 - 1200 nm wavelength range reveals interesting features. Peaks approximately at 683, 807, 883 and 980 nm are observed for both samples with and without the SiO$_2$ layer (better visualized in the inset of figure 2). The peak at ~ 926 nm is only observed for the sample with the SiO$_2$ layer. Therefore this peak is most probably related to the multilayer resonance, particularly, to the calculated resonance near 950 nm as describe above On the other hand, the peaks at ~ 683 and ~ 883 nm can be attributed to radiative recombination at localized tail states Their energy positions depend on the x fraction of oxygen in the $a$-SiO$_x$ matrix [18]. Also it has been reported the possibility that $a$-Si-rich clusters and $a$-SiO$_2$ grains existent in an $a$-SiO$_x$ matrix may cause the emission at these wavelengths [18]. The fact that Q-factor is small leads to some enhancement of the emission at 883 nm emission. Finally, the peaks at ~ 807 and ~ 980 nm may in principle be assigned to the intrinsic Er 4$f$ shell emission ($^4I_{9/2}$ to $^4I_{15/2}$ and $^4I_{11/2}$ to $^4I_{15/2}$, respectively). In particular, the spectral feature at 980 nm was also observed in a similar matrix [9]. However, care must be taken before PL at ~ 807 and ~ 980 nm can definitively be attributed to Er$^{3+}$. This issue will be discussed later.

We proceed to investigate further improvements in the emission efficiency with the use of thermal treatment on the samples with the SiO$_2$ resonant structure. Figure 2 shows the PL spectra taken at 295 K in the 600 - 1200 nm wavelength range after annealing for 1 hour at 400 °C and at 1150 °C. We kept the spectra for the as-deposited on Si sample fo comparison. After annealing at 400 °C there is over 8 times increase in PL intensity with respect to the as-deposited sample In addition, the emission spectral line-shape and the peaks position characteristic of the as-deposited sample remain apparently not affected upon this thermal treatment. On the other hand, after annealing at 1150 °C, the PL intensity shows expressive reduction. Moreover, the PL line-shape and feature peaks change considerably. It is worth to observe that afte annealing at 1150 °C, the peak at ~ 980 nm is totally suppressed while the peak at ~ 807 nm remains. Upon high temperature annealing, silicon segregates forming clusters and there is great healing of the defects generated by the oxygen vacancy dangling bonds. Therefore, we expect that most of the defect-related processes, such as the Er$^{3+}$ resonan pumping and the defect-related emission, should be greatly decreased. This can explain the large overall reduction of the PL. Furthermore, we believe the 980 nm peak arises from the Er $^4I_{11/2}$ - $^4I_{15/2}$ transition and the 807 nm peak is probably generated by some localized radiative defects and coincidently have similar energy than the $^4I_{9/2}$ - $^4I_{15/2}$ transition.

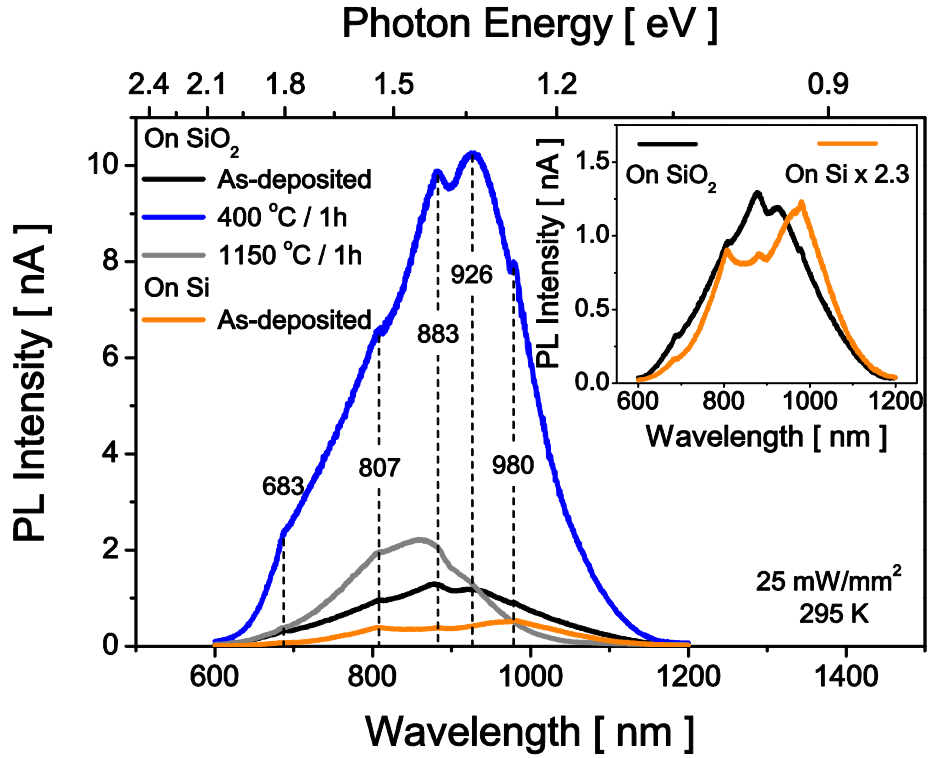

**Figure 2**. PL spectra at 295 K in the visible - near infrared region (600 - 1200 nm) for the samples: as-deposited on Si, as-deposited on $SiO_2$ and submitted to two different thermal treatment temperatures. The inset shows a 2 fold enhancement of PL intensity due to the resonant layer.

Figure 3 shows the PL spectra in the 1400 - 1700 nm wavelength range for the same samples shown in figure 2. The main peak (highest-intensity at ~ 1540 nm) corresponds to the $^4I_{13/2}$ - $^4I_{15/2}$ transition and the second peak at ~ 1550 nm is caused by the Stark effect [12]. The spectra corresponding to the as-deposited on $SiO_2$ sample also present a substantial enhancement in the emission (about 4 times - see inset of figure 3) with respect to the as-deposited on Si sample. This enhancement in the PL caused by the multilayer structure can be qualitatively explained considering a system with internal spontaneous/stimulated emission generation coupled to several modes with different cavity Q's. Using a simple model based on rate equations the emission power out of the system at a given mode $\omega_i$ can be given by:

$$R_i = \frac{\beta_i R_{sp} \hbar \omega_i}{(A-B)Q_{ci}/\omega_i + 1} \quad (2)$$

where $A$ and $B$ are the scattering loss and the stimulated emission rates per photon, respectively. $Q_{ci}$ and $\beta_i$ are the cold cavity Q and the spontaneous emission factor for the $i^{th}$ mode. $R_{sp}$ is the total spontaneous emission rate. At transparency, $A = B$, the emission is not affected by the different Q's and only modulated by $R_{sp}$ and $\beta_i$. For $A < B$, modes with smaller cold cavity Q should have higher emission power. Essentially, lower Q means light generated within the system is not confined and is readily emitted. On the other hand, only if $A > B$, the higher cold cavity Q emission should be enhanced. On other words, in this last case, the higher cold cavity Q allows a photonic density build up that compensates the higher confinement. This explains the two-fold higher emission near 926 nm for the samples with the $SiO_2$ layer only if stimulated emission occurs. The stimulated emission may occur for the defect-related emission and/or for the $Er^{3+}$ transition at 980 nm ($^4I_{11/2}$ - $^4I_{15/2}$). However, since $^4I_{11/2}$ quickly decays to $^4I_{13/2}$ level there could be a cross-pumping at 1540 nm. Both the large mode overlap and the low Q at this wavelength lead to a large photon escape from the cavity. In summary, the larger than expected enhancement of the emission at 1540 nm may be explained by the low Q at this wavelength, allowing high external quantum efficiency combined with a higher internal emission rate at this wavelength from the more confined resonance at 980 nm.

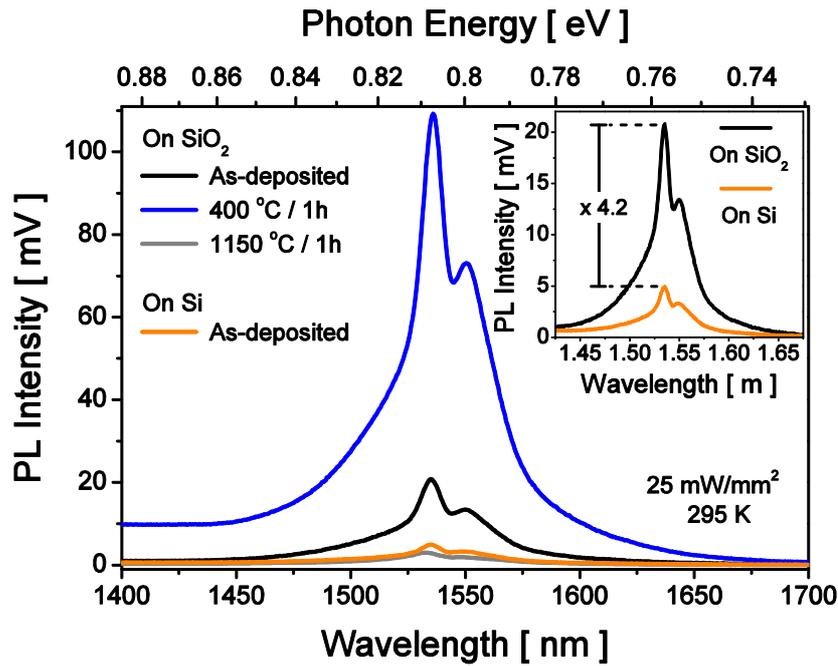

**Figure 3**. PL spectra at 295 K in the near-infrared region (1400 - 1700 nm) for the samples: as-deposited on Si, as-deposited on SiO$_2$ and submitted to thermal annealing at 400 °C and at 1150 °C for 1 hour. The inset shows a PL enhancement of over 4 times due to the resonant layer.

We also observed a PL enhancement in the near-infrared (NIR) region of over 5 times after thermal treatment of the as-deposited on SiO$_2$ sample at 400 °C / 1h. This result suggests that a larger fraction of the Er atoms reacts with oxygen during annealing (Er oxidation) forming a greater amount of Er-O optically active complexes [19]. Moreover, it may suggest that there is a higher effective defects density (dangling bonds) that acts as Er luminescence sensitizers. It was showed that any defect due to excess Si in silicon oxide may act as a sensitizer [16]. On the other hand, high temperature annealing leads to a drastic deterioration in the NIR emission efficiency, that is, the Er$^{3+}$ pumping was highly suppressed. The severe decrease observed in PL intensity could be understood as the contribution of two effects: *i*) decrease of the Er$^{3+}$ excitation rate, and/or *ii*) lower density of optically active Er$^{3+}$ centers as a result of the local and surrounding environment of the Er ions.

The decrease of the Er$^{3+}$ excitation rate is due to a reduction of the density of sensitizers with the increase of temperature. This reduction can be detected by PL spectra obtained in the visible - near infrared range (figure 2), where there is a drastic reduction of emission from *a*-SiO$_x$ defect-related radiative centers after annealing at 1150 °C - considering that either the excitation of these centers is mediated by dangling bonds or they themselves emit. Accordingly, the annealing at high temperature should produce a rearrangement and/or a partial annihilation of the dangling bonds leading to a decrease of the Er$^{3+}$ excitation rate too. Indeed, we notice that the PL [1150 °C]/[400 °C] ratio at 600 - 1200 nm range is about 0.22, while at the 1400 - 1700 nm range is about 0.027. This indicates that there is not only a decrease in the density of the sensitizers, but also a decrease in the number of optically active Er$^{3+}$ centers. It is known that the PL intensity depends crucially on the Er incorporation conditions on the host matrix, since the Er$^{3+}$ transition probability is determined by the local environment of the ions - more specifically determined by the magnitude of the crystal field which depends on the Er$^{3+}$ sites symmetry (centrosymmetric or non-centrosymmetric lattices) [20]. The luminescence efficiency has been associated to an Er local environment very similar to that of Er$_2$O$_3$ where the Er sits in a 6-fold coordinated cage (each Er is coordinated by 6 oxygen atoms), which provides a non-centrosymmetric environment [21,22]. Tessler *et al*. investigated the possible Er lattice sites and their evolution upon thermal annealing (up to 1100 °C) in samples with an amount of oxygen atoms similar to ours [19,23]. They observed that the erbium first neighbor shell consists always of oxygen atoms and does not change under annealing. The erbium chemical environment evolves with temperature from a characteristic ErO$_3$ 3-fold coordinated oxygen shell towards the Er$_2$O$_3$-like 6-fold coordinated shell (where the 3-fold Er coordination has a lower symmetry than the 6-fold coordination). Although this increase of coordination with annealing temperature (the Er$^{3+}$ sites become more symmetric) gives rise to a decrease of the transition probability, the main consequence is Er ions surrounding structural change. It was reported that in Er-doped silicon-rich silicon oxide matrix containing Si-NC's, Er$^{3+}$ can exist in two types of centers: isolated or strongly coupled to defects [16]. For the case of isolated centers the emission efficiency of Er$^{3+}$ is significantly limited by distance-dependent energy transfer, which corroborates the results at higher temperature.

In order to make a more meaningful distinction between the types of centers abovementioned, we have investigated the variation of emission intensity of the Er-O complexes with temperature. The temperature dependence of the integrated PL intensity for the annealed samples at 400 °C and at 1150 °C (from now on labeled as S1 and S2, respectively) as a function of the inverse temperature (1/k$_B$T) is depicted in figure 4. The data were taken at temperatures ranging from 10 to 295 K at a temperature were observed. The PL intensity declines more rapidly up to 70 K for the S2 sample, above which a much stronger quenching for both sample occurs.

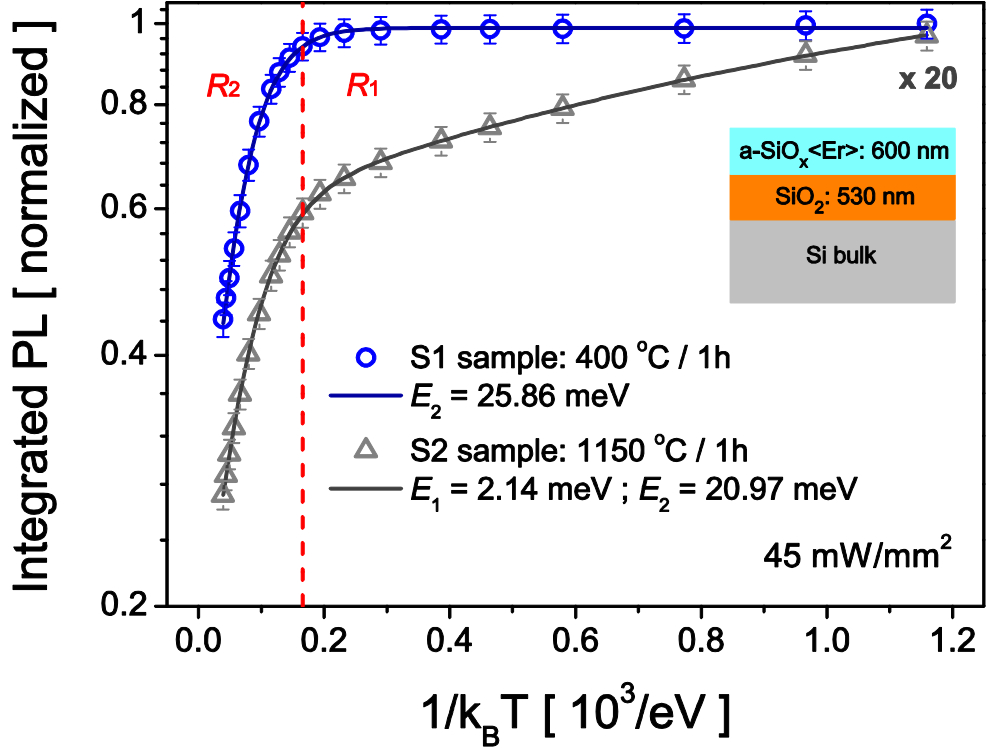

**Figure 4**. Arrhenius plot of the integrated PL intensity as a function of $1/k_BT$. The symbols represent the experimental data, while the continuous lines are the theoretical fits according to equation (3).

Based on the model of the non-radiative recombination, the thermal quenching properties were quantified and analyzed assuming characteristics thermally deactivated excitation processes, where the temperature dependence of integrated PL intensity $I_{PL}(T)$ is given by [24]:

$$I_{PL}(T) = I_0 / \left[1 + \sum_{i=1}^{n} C_i \exp(-E_i / k_B T)\right] \quad (3)$$

where $C_i$ are the coefficients associated to the thermal dissociation processes of Er centers with deactivation energy $E_i$, $I_0$ is the PL intensity at a temperature close to absolute zero and $k_B$ is the Boltzmann constant. $I_{PL}(T)$ is obtained by fitting Gaussian function on the main peak at ~ 1540 nm. The fitting results of the experimental data according to equation (3) are represented in figure 4 by continuous lines. For the S1 sample the quenching process is well described by only one regime (solely a deactivation energy $E_2$ ~ 26 meV) while that for the S2 sample two quenching regimes (with deactivation energy $E_1$ ~ 2 meV and $E_2$ ~ 21 meV) are clearly observed. The fitting parameters used to model temperature-dependent Er-PL behavior are presented in table I, as well as the ratio $R$ of the integrated PL intensity for two quenching ranges: $R_1 = [I_{PL} - 70$ K$] / [I_{PL} - 10$ K$]$ and $R_2 = [I_{PL} - 295$ K$] / [I_{PL} - 70$ K$]$.

**Table I**. PL parameters obtained from temperature dependence of the integrated PL intensities.

| Sample | $C_1$ | $E_1$ (meV) | $C_2$ | $E_2$ (meV) | $R_1$ | $R_2$ |
|---|---|---|---|---|---|---|
| S1 - 400 °C / 1h | - | - | 3.7 | 25.86 | 0.94 | 0.47 |
| S2 - 1150 °C / 1h | 1.18 | 2.14 | 5.83 | 20.97 | 0.61 | 0.46 |

Supposing that the thermal ionization of [Er center - dangling bond defect] pair before the electronic excitation transfer process to occurs is dominant on backtransfer mechanism (multi-phonon non-radiatives processes) [7], it is reasonable to assume that such energies are associated to Er center - dangling bond decoupling. Remembering that for an effective energy transfer to the 4 $f$ electrons either by DRAE process (capture of an electron by dangling bonds) or by resonan dipole-dipole interaction originated by the non-radiative recombination of the electron-hole pairs in the dangling bonds, it is necessary that the dangling bond be localized nearby the Er-O complexes. Hence, the integrated PL intensity is the responses average from Er-O complexes and their vicinity. At lower temperatures (T < 70 K), the thermalization energies and the quenching rate values (table 1) suggest that the deactivated centers of S1 sample differ completely from the S2 sample. Regarding to the latter, possibly photogenerated carriers near weakly coupled centers have less kinetic energy to out diffuse (are more localized) allowing the excitation. In contrast, as temperature increases, the carriers become more delocalized and the quench is very effective for those centers. At higher temperatures (T > 70 K) the Er centers for the two samples differ slightly (deactivation energies and quenching ratio similar) apparently due to a small difference in the local order and surrounding of the Er atoms. However, the integrated PL intensity of S2 sample is about 97 % lower than S1 sample, which reveals a lower density of optically active $Er^{3+}$ centers. At this point, we also speculate that the annealing at higher temperatures induced some matrix phase separation, or even further Er precipitation that may promote the reduction of active centers.

## 4. Conclusions

In summary, we have fabricated highly-luminescent samples with erbium-doped amorphous silicon sub-oxide (*a*-SiO$_x$<Er>) layers deposited on $SiO_2$/Si substrates forming a resonator structure. Two-fold improvement in PL intensity of the *a*-SiO defect-related radiative centers is achieved in the wavelength range between 800 - 1000 nm due to the resonator structure. The PL intensity in the wavelength range between 1400 - 1700 nm (region of $Er^{3+}$ $^4I_{13/2}$ - $^4I_{15/2}$ transition) is increased four times apparently due to optical cross-pumping at 980 nm ($^4I_{11/2}$ - $^4I_{15/2}$ transition). After temperature annealing optimization the emission at 1540 nm was further enhanced over 5 times as compared to the as-deposited sample.

## Acknowledgements

This work was partially supported by the Brazilian financial agencies CNPq and FAPESP through the National Institute for Science and Technology in Optics and Photonics (FOTONICOM).